\documentclass[aps,prb,twocolumn,groupedaddress,amsmath,amssymb]{revtex4}

\usepackage{graphicx}
\usepackage{dcolumn}
\usepackage{bm}

\graphicspath{{../TheEnd/CH6/figures/}}

\newcommand{\Tcdw}[0]{T$_{\textrm{CDW}}$ }
\newcommand{\cmm}[0]{cm$^{-1}$ }
\newcommand{\ud}{\mathrm{d}}

\begin{document}

\title{Evidence for strongly coupled charge-density-wave ordering in three-dimensional RE$_5$Ir$_4$Si$_{10}$ compounds by optical measurements}

\author{Riccardo Tediosi\footnote{Current address: Bruker Biospin AG, Industriestrasse 26, Ch-8117 Fallanden, Switzerland}}
\author{Fabrizio Carbone\footnote{Current address: Arthur Amos Noyes Laboratory of Chemical Physics
California Institute of Technology, Mail Code 127-72, 1200 East California Boulevard, California, USA}}
\author{A. B. Kuzmenko}
\author{J. Teyssier}
\author{D. van der Marel}
\affiliation{DPMC, University of Geneva, 24, Quai Ernest-Ansermet,
1211 Geneva 4, Switzerland}
\author{J. A. Mydosh}
\affiliation{Institute of Physics II, University of Cologne, D-50937 Cologne, Germany}

\date{March 16, 2009}

\begin{abstract}
We report optical spectra of Lu$_5$Ir$_4$Si$_{10}$ and Er$_5$Ir$_4$Si$_{10}$, exhibiting the phenomenon of coexisting superconductivity or antiferromagnetism and charge density wave (CDW) order. We measure the maximum value of the charge density wave gap present on part of the Fermi surface of Lu$_5$Ir$_4$Si$_{10}$, corresponding to a ratio $2\Delta/k_BT_{\textrm{CDW}}\simeq 10$, well above the value in the limit of weak electron-phonon coupling. Strong electron-phonon coupling was confirmed by analyzing the optical conductivity with the Holstein model describing the electron-phonon interactions, indicating the coupling to phonons centered at 30 meV, with a coupling constant $\lambda \approx 2.6$. Finally we provide evidence that approximately 16\% of the Fermi surface of Lu$_5$Ir$_4$Si$_{10}$ becomes gapped in the CDW state.
\end{abstract}


\maketitle
\section{\label{intro}Introduction}
Charge-density-wave (CDW) formation was demonstrated \cite{peierls} to be the consequence of an interaction between phonons connecting nested sides of the Fermi surface (FS) and metallic electrons in the system. For a particular phonon mode with wavevector $Q = 2k_{F}$ a charge density modulation is induced leading to a rearrangement of the original atom positions which tend to open a gap at the FS. When the temperature of the system decreases below the transition temperature $T_{\textrm{CDW}}$, a full gap can be ideally opened and the system can undergo a metal-insulator transition (MIT), also known as Peierls transition \cite{peierls,gruner}. Experimentally the density wave formation was demonstrated in many quasi-monodimensional compounds like K$_{0.3}$MoO$_{3}$ \cite{travaglini84,travaglini81,degiorgi91}, also known as ``blue bronze'', NbSe$_{3}$ \cite{perucchi04,fournel86,sridhar85}, TaSe$_{3}$ \cite{sridhar85} and in different families of organic materials (TTF-TCNQ and Bechgaard salts \cite{schwartz98,henderson99,giamarchi04}).

For weak electron-phonon interactions, also known as the weak coupling regime, the relation between the value of the gap and the mean-field CDW transition temperature $T^{\textrm{MF}}_{\textrm{CDW}}$ obeys the relation $2\Delta = 3.52 k_B T^{\textrm{MF}}_{\textrm{CDW}}$ well known for the case of BCS superconductivity. In the situation where the weak electron-phonon coupling condition is not fulfilled, deviations from the standard BCS-like scenario are expected. For a purely monodimensional system long range order develops at the transition temperature $T_{\textrm{CDW}}^{\textrm{MF}}$; the ground state is characterized by a coherence length $\xi_0$ that represents the spatial dimension of the coupled e-h pair.

McMillan \cite{mcmillan77} proposed a description beyond
the weak-coupling limit. He pointed out that, since in the weak coupling regime the phonon frequencies are modified over a small region of the reciprocal space, the phonon entropy is unimportant with respect to the electrons' entropy.
In the opposite case of a short coherence length,
the phonon contribution to entropy dominates over
the electron's contribution. In the strong coupling regime the ratio $2\Delta/k_B T_{\textrm{CDW}}$ exceeds the weak-coupling limit by a factor of 2 for realistic values of the electron-phonon coupling. This theory has been successfully applied to practical cases where a CDW transition was accompanied by anomalously large changes in specific heat, thermal conductivity as well as thermoelectric power. For example it could explain the anomalies observed in $2H$-TaSe$_{2}$ \cite{vescoli98,mcmillan77}.

In this regard Lu$_{5}$Ir$_{4}$Si$_{10}$ has many interesting properties that deviate
from the standard weakly coupled CDW behavior and shows signatures of a
strongly coupled system\cite{ramakrishnan07,becker99}. In this compound the CDW and the
superconducting transitions were found at
T$_{\textrm{CDW}}=83$ K and T$_{\textrm{SC}}=3.9$ K and the two
states coexist below the superconducting transition as confirmed via
X-ray-diffraction \cite{becker99}. The first intriguing
aspect is that the anisotropy of the material
($\rho_{a}/\rho_{c}\simeq3.5$) is much lower than expected for
typical monodimensional systems. Furthermore the system does not go
through a pure metal-insulator transition across $T_{\textrm{CDW}}$. It remains
metallic down to the superconducting transition where eventually the
resistivity drops to zero. The other interesting aspect concerns its
thermodynamical properties, which show the anomalies characteristic of strongly coupled systems. The specific heat, thermal
conductivity and thermoelectric power
measurements \cite{becker99,jung03,kuo01}, indicate a large
specific-heat jump ($\Delta C_{p}=55$ J/mol K for a polycrystalline
sample \cite{kuo01} and $\Delta C_{p}=160$ J/mol K for a single-crystal
sample \cite{becker99}) developing in a very narrow temperature range
($\Delta T_{\textrm{CDW}}\sim 1\%$) accompanied by a significant entropy
variation ($\Delta S=0.12R$\cite{kuo01}, $\Delta S=0.5R$\cite{becker99}, where $R = $ gas constant). The relative size of the specific heat variation has been compared to the expected $1.43$ value predicted assuming a weak coupling approximation; the result showed an important deviation from this latter value giving $\Delta C_e/C_e \simeq 8.4$, where $C_e$ represents the  electronic contribution to the specific heat\cite{kuo01}. All these features imply strong coupling\cite{mcmillan77}.

The isostructural compound Er$_{5}$Ir$_{4}$Si$_{10}$ on the other hand shows two CDW transitions: the first one at 155 K is a transition to an incommensurate CDW state while at 55 K another transition establishes the final commensurate CDW state. Also in this case the CDW instability originates apparently from a 3D system, since the resistivity ratio along the two main principal axis $\rho_a/\rho_c \sim 2.33$. The substitution of lutetium with erbium adds a net magnetic moment, which creates at low temperature ($2.8$ K) an anti-ferromagnetically ordered state without displaying any signature of superconductivity as observed for the Lu$_{5}$Ir$_{4}$Si$_{10}$ compound.
Also in this case it has been shown that the charge-density-wave state and the antiferromagnetic ordering coexist below the Neel temperature.

In this paper we report a study of the optical properties of Lu$_{5}$Ir$_{4}$Si$_{10}$ and Er$_{5}$Ir$_{4}$Si$_{10}$
single crystals in the temperature range between 20 K and room
temperature and in the spectral region between 0.01 and 4 eV. The optical spectra of the two compounds are almost identical, but the detailed temperature dependence is different due to differences in the manner in which the CDW order evolves as a function of temperature in the two materials. For both materials the plasma frequency
along the main optical axes turns out to be relatively isotropic, around $1.13$ for the ratio
$\omega^{\ast}_{Pc}/\omega^{\ast}_{Pa}$. Detailed analysis of the CDW gap and the spectral weight of Lu$_{5}$Ir$_{4}$Si$_{10}$ reveals that about 16\% of the Fermi surface becomes gapped, and that the CDW phenomenon is in the strong coupling limit of McMillan's theory of electron-phonon coupling in CDW systems.

\section{\label{exp}Experimental methods}
\begin{figure}
\includegraphics[width=0.48\textwidth]{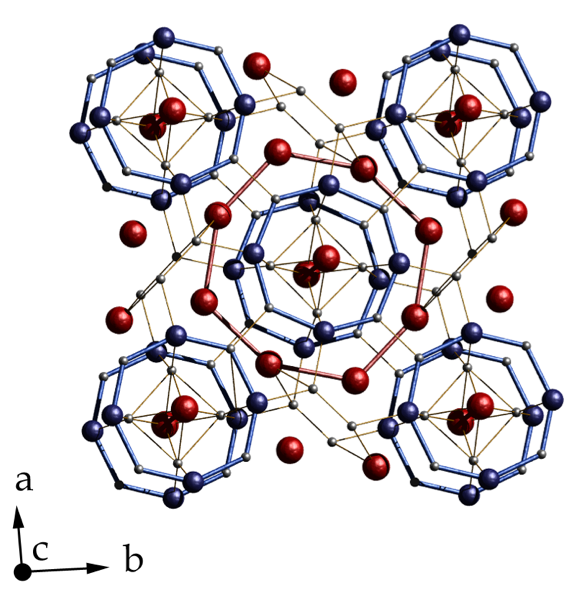}
\caption{\label{fig:crystal_structure} (color online) The crystal structure of RE$_5$Ir$_4$Si$_{10}$. The rare earth atoms
are represented with large (red) spheres, Ir atoms with medium (blue)
spheres and Si atoms with small (light-grey) spheres. With thick (blue) lines
are represented schematically the five chains extending along the
\emph{c}-axis which seems to be responsible for the quasi-monodimensional
behavior of the system and with thick (red) lines the RE-RE network
separating the five chains.}
\end{figure}
RE$_5$Ir$_4$Si$_{10}$ single-crystals were grown using a modified Czochralski
technique as described in Refs.~\cite{becker99,menovsky}. Its
crystal structure belongs to the tetragonal
Sc$_{5}$Co$_{4}$Si$_{10}$ family (\emph{P4/mbm}), thus the \emph{a}
and \emph{b}-axis are degenerate $a=b=12.4936$ ${\AA}$, while the
\emph{c}-axis size is $c=4.1852$ ${\AA}$. The atom positions are
represented in Fig.~\ref{fig:crystal_structure}; the dark (blue) lines are the
Ir-Si-Ir bonds around the RE atoms placed in the corners and at the
center of the main cell, while the thick (red) lines are the RE-RE bonds.
The five RE atoms seated at the corners and in the center of the
cell surrounded by four Ir atoms interleaved with Si atoms are separated by a network of eight RE atoms in the
\emph{ab} plane. This description \cite{becker99} was introduced to
explain the two-fold nature of this class of compounds: on the one hand the CDW
transition is a signature of monodimensionality, while on the other hand a constant
metallic behavior below $T_{\textrm{CDW}}$ indicates that the transport
perpendicular to the chains is not completely forbidden. By analyzing
the distances between atoms in the main cell it was found that RE atoms
in the corners and in the center of the unit cell have the shortest
interatomic distance with respect to all the other bonds; this suggests that the monodimensional conducting channel is the RE-RE chain developing along the \emph{c}-axis (perpendicular to the paper sheet). Accordingly, the RE network around the central site can be viewed as a channel of conductivity perpendicular to the chains.

The crystals (surface area $\sim 18$ mm$^{2}$ and thickness $\sim 1$ mm) were oriented with
a Laue camera, and polished with diamond sand paper of 0.1 $\mu$m
grain size until a flat and mirror-like quality surface was
obtained. The samples, whose optically treated surface contained the
\emph{ac} plane of the crystal, were mounted on a conical copper
holder using silver paint to guarantee an optimal thermal contact,
and aligned with respect to the optical spectrometers using a He-Ne laser
source.

\begin{figure}
\includegraphics[keepaspectratio,width=0.48\textwidth]{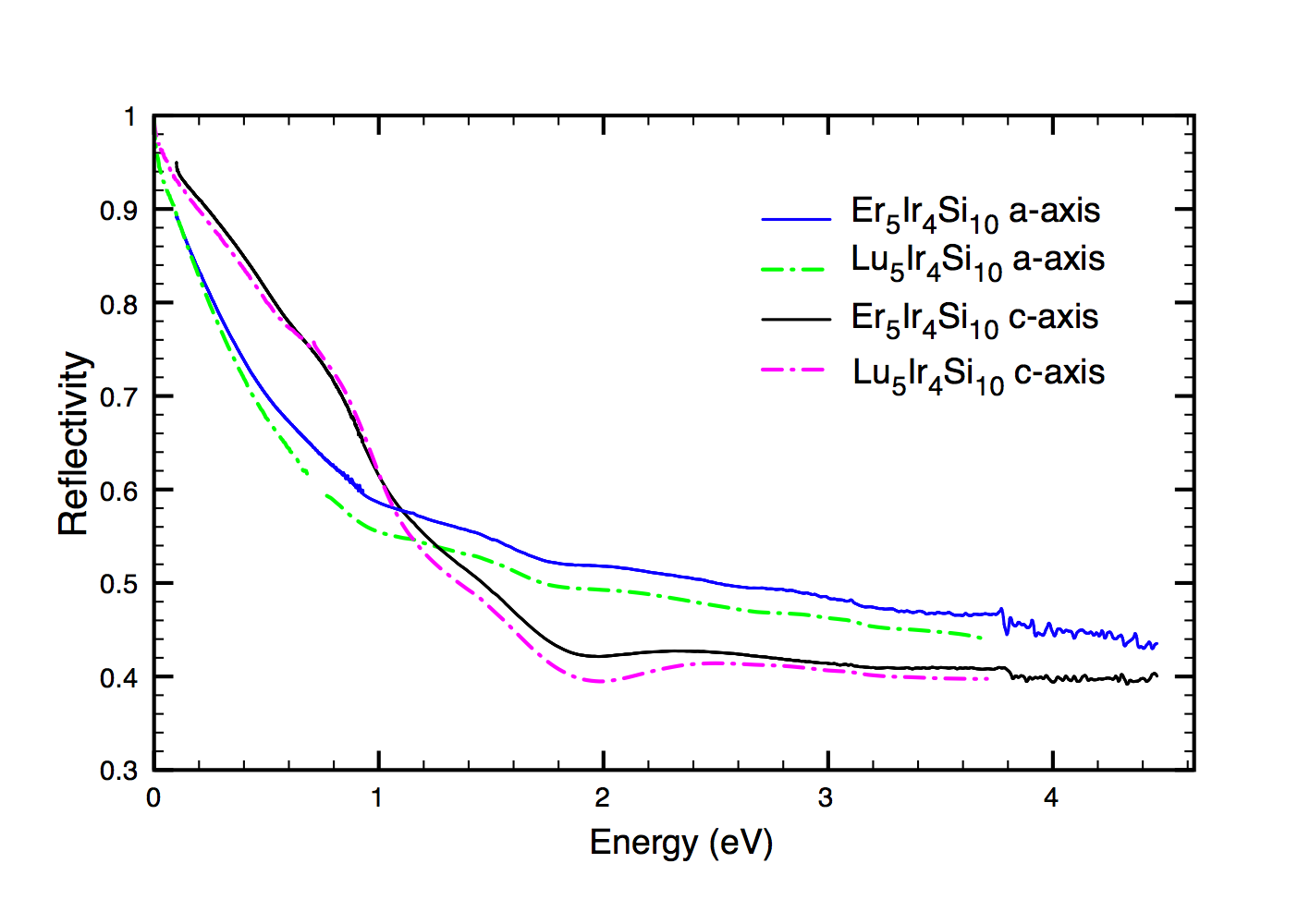}
\caption{\label{fig:IR_Ref} (Color online) Optical reflectivity with $a-$ and $c$-axis polarized light of Lu$_{5}$Ir$_{4}$Si$_{10}$ (dashed curves:  green for $a$-axis and magenta for $c$-axis) and of Er$_{5}$Ir$_{4}$Si$_{10}$ (solid curves: blue for $a$-axis and black for $c$-axis). The spectra are quite similar for both materials. The reflectivity in the infrared range is much higher along the $c$-axis than along the $a$-axis for both materials, confirming the more metallic characteristics along the $c$-axis in this class of materials. }
\end{figure}
\begin{figure}
\includegraphics[keepaspectratio,clip,width=0.48\textwidth]{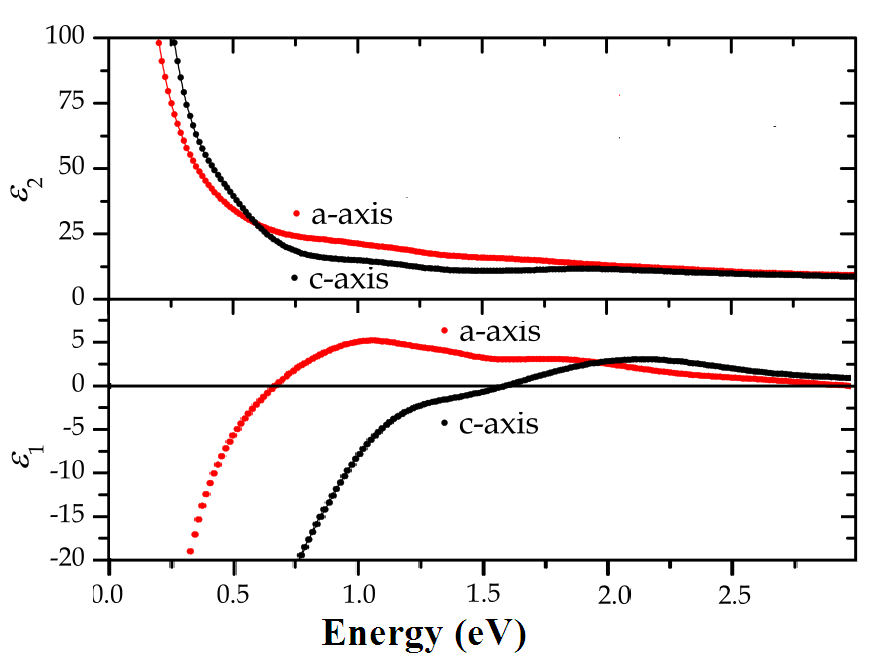}
\caption{\label{Lu_epsilon_room} (Color online)  The real (bottom) and imaginary (top) parts of the dielectric function of Lu$_{5}$Ir$_{4}$Si$_{10}$. The strong anisotropy in the screened plasma frequency $\varepsilon_1(\omega^*_p)=0$, observed for $E\parallel a$ (red circles) and $E\parallel c$, reveals a larger screening effect along \emph{a}-axis due to interband contributions.}
\end{figure}

The optical reflection of Lu$_{5}$Ir$_{4}$Si$_{10}$ and Er$_{5}$Ir$_{4}$Si$_{10}$ were measured in the wavelength range from
50 \cmm (6.2 meV) up to 30000 \cmm (3.8 eV) using two different
techniques: Fourier-transform spectroscopy in the
infrared range and ellipsometry for the visible and UV region. In
the infrared ($6.2\rightarrow 700$ meV) we used a custom-made
optical cryostat, whose stability allowed reflected signal variations on the order of $0.1\%$ to be measured, in combination with a
Bruker 113 spectrometer. The samples were mounted in a quasi-normal
incidence configuration and the absolute value of the reflectivity
$R(\omega)$ was calculated by taking as a reference measurement a gold
layer evaporated \emph{in situ}. We acquired data continuously
during slow temperature sweeps ranging from 295 K down to 20 K (1 K
of resolution) and we used a rotating polarizer in order to measure
the response of the \emph{a}- and \emph{c}-axis orientations. Figure \ref{fig:IR_Ref} shows the experimental reflectivities for the two samples along the \emph{a-} and \emph{c}-axis. The comparison reveals large similarities in the infrared and visible response of the two materials. In both cases the \emph{c}-axis has a more metallic response below 1.3 eV in agreement with its smaller dc resistivity \cite{becker99}. The much higher infrared reflectivity along the c-axis confirms that the chains parallel to the \emph{c}-axis (Fig.~\ref{fig:crystal_structure}) are  the preferred conductivity channels in this class of compounds. The $a$ and $c$-axis reflectivity curves cross at 1 eV. The lower c-axis reflectivity above this energy is caused by the higher value of Im $\epsilon(\omega)$ along the $a$-axis, as shown in Fig. \ref{Lu_epsilon_room}. The different values of Im $\epsilon(\omega)$ along the $a$ and $c$ axis reflect the anisotropy in the optical matrix-elements of the interband transitions.

In the ellipsometry measurements we used an angle of incidence of
74$^{\circ}$ with respect to the normal to the sample surface. A UHV cryostat with a base pressure of
$10^{-9}$ mbar was used, resulting in no noticeable absorption of gas on the sample surface during cooling. The \emph{a}- and \emph{c}-axis responses
were obtained rotating the sample in the cryostat. The alignment of
the crystal axis with respect to the incoming polarized light was carefully checked
in order to reduce the alignment error to values smaller than 1$^{\circ}$. The screened plasma energies $\hbar\omega^*_p$, defined as $\varepsilon_1(\omega^*_p)=0$, are directly visible in the ellipsometric data (Fig.~\ref{Lu_epsilon_room}). The plasma energy is strongly anisotropic: 0.69 eV along \emph{a} and 1.61 eV along \emph{c}. This anisotropy is due to a stronger screening by the interband transitions along the \emph{a}-axis and does not reflect the real anisotropy of the free carrier spectral weight(see section \ref{section 7})

\begin{figure}
\includegraphics[keepaspectratio,clip,width=0.48\textwidth]{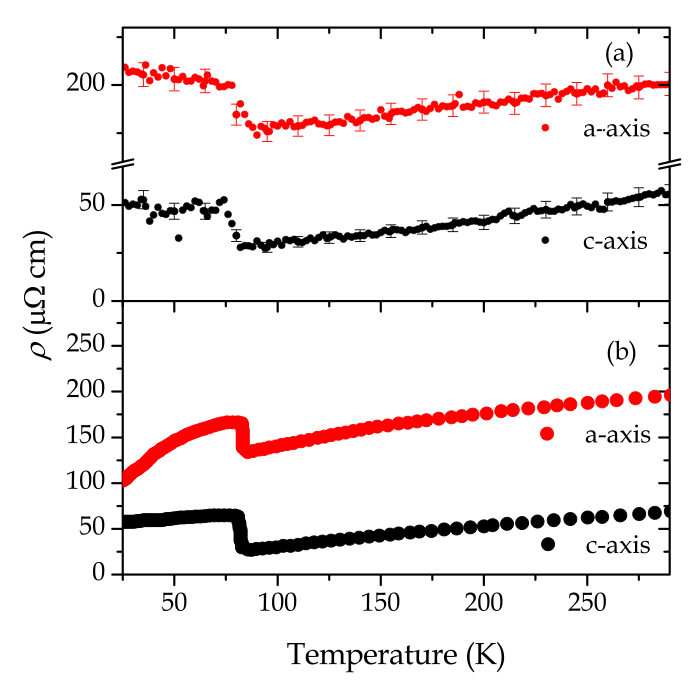}
\caption{\label{resistivity_comparison} (Color online) (a) DC resistivity extrapolated from
optical data as a function of temperature for \emph{a}- (red circles)
and \emph{c}-axis (black circles). The error bars are calculated
considering an error in the absolute value of reflectivity of $\pm
0.5$ \%. (b) Four terminal resistivity from Ref.~\onlinecite{becker99}.}
\end{figure}
In order to obtain the real and imaginary part of the optical conductivity along the $a$- and the $c$-axis in the entire frequency range, we used a variational routine yielding the Kramers-Kronig consistent dielectric function which reproduces all the fine details of the infrared reflectivity spectra while simultaneously fitting to the ellipsometrically measured complex dielectric function in the visible and UV-range\cite{alexey}. In contrast to the "conventional" Kramers-Kronig transformation of the reflectivity spectra, this procedure anchors the phase of the reflectivity to the phase in the visible range measured directly with ellipsometry\cite{bozovic90}. To check the reliability of our data analysis, we compare in Fig. \ref{resistivity_comparison} the inverse optical conductivity extrapolated to zero frequency, $\sigma_1(0,T)^{-1}$ with the dc resistivity measured on the same sample using the 4 terminal method\cite{becker99}. Upon reaching the lowest temperature, the optical \emph{a}-axis $\sigma_1(0,T)^{-1}$ departs progressively (up to a factor 2) from the four-probe $\rho_{dc}$. This discrepancy finds a natural explanation in the observation that, along the $a-axis$, the Drude peak is weak compared to the interband transitions (see section \ref{section_T_dependence}). When we lower the temperature below $T_{CDW}$, the Drude-peak narrows, and the Drude optical conductivity decreases in the measured frequency range ($\hbar\omega > $ 6 meV). Consequently the extrapolation of the $a$-axis optical conductivity below 6 meV (for which we use the Drude model) becomes less reliable at low temperatures.

\section{Temperature dependence of the optical constants} \label{section_T_dependence}
\paragraph{Er$_{5}$Ir$_{4}$Si$_{10}$}\label{Er_sub}
\begin{figure}
\includegraphics[keepaspectratio,clip,width=0.48\textwidth]{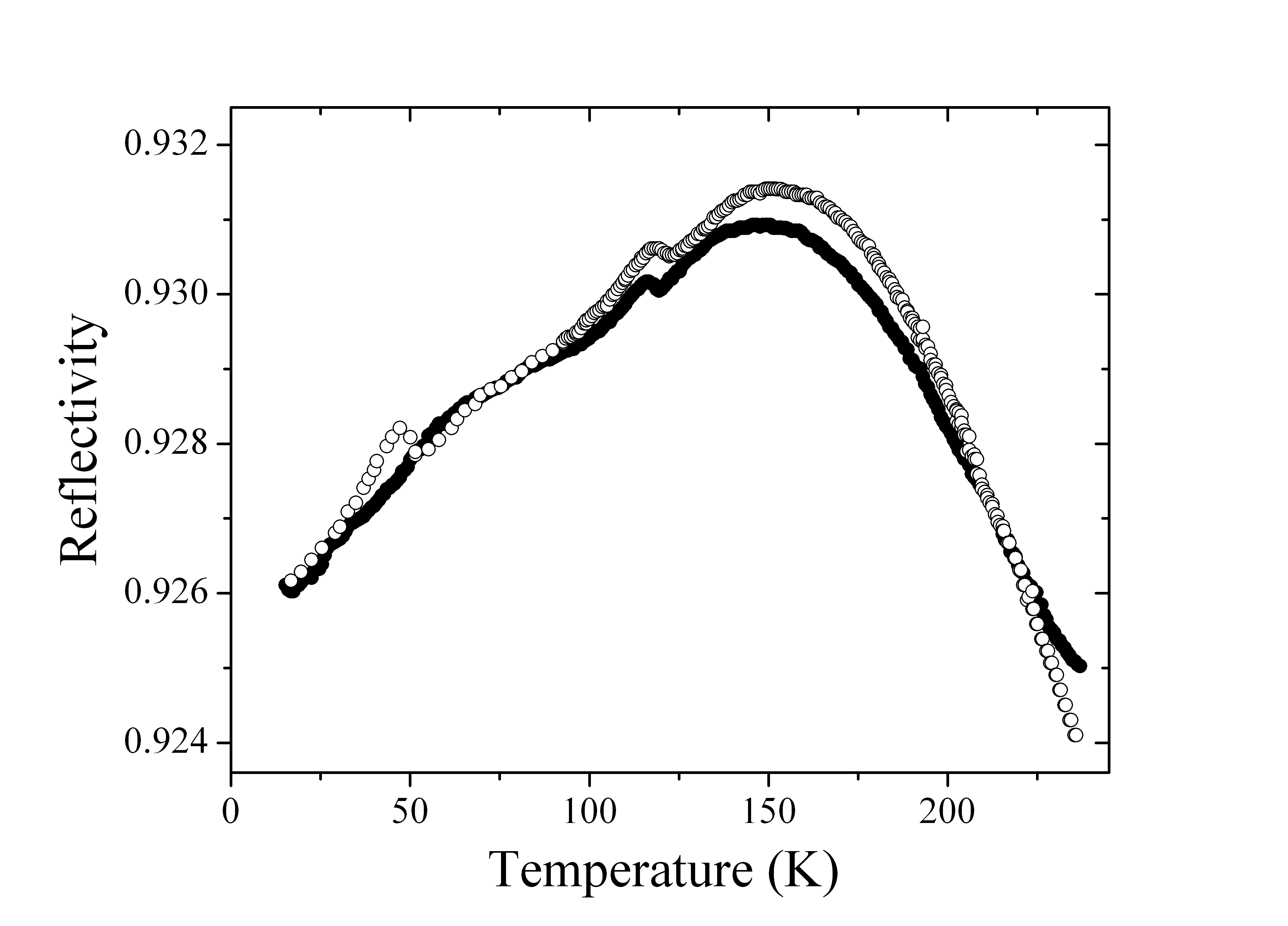}
\caption{\label{ErIrSi_signal} Reflectivity of Er$_{5}$Ir$_{4}$Si$_{10}$ at 186 meV measured along the \emph{c}-axis during cooling (closed symbols) and heating (open symbols). The measurement shows clearly the two CDW transitions at 148 and 53 K. The low temperature CDW transition at 53 K is characterized by a reflectivity change of only $0.05\%$. The hysteresis for this transition while cooling and heating the sample testifies to the first order nature of this phase transition.}
\end{figure}
As noted in Sec.~\ref{intro}, the Er based compound should show two separate transitions at two different temperatures. The temperature dependence of the reflected signal at 186 meV, shown in Fig.~\ref{ErIrSi_signal}, reveals these two transitions at 148 and 53 K, consistent with previously published transport data \cite{galli2000PRL,jung03}. The low temperature transition shows a clear temperature hysteresis, consistent with the transport results. The absolute change of the signal at the two phase transitions is only of the order of $0.05\%$ for light polarized along the $c$-direction, wile the CDW transition was not observable at all in the $a$-axis optical spectra of Er$_{5}$Ir$_{4}$Si$_{10}$. The CDW induced changes in reflectivity are hence a factor 10 weaker than in Lu$_{5}$Ir$_{4}$Si$_{10}$. Also the DC resistivity has a completely different behavior\cite{jung03}, showing a weak (about 5\%) stepwise increase below 148 K, and a gradual decrease below 53 K. Whereas Lu$_{5}$Ir$_{4}$Si$_{10}$ transforms directly from the metallic phase to a CDW at 83 K, in Er$_{5}$Ir$_{4}$Si$_{10}$ these two phases are separated by an incommensurate CDW between 55 and 155 K. The two phase transitions in Er$_{5}$Ir$_{4}$Si$_{10}$ are therefore of a qualitatively different nature than the metal to CDW transition in Lu$_{5}$Ir$_{4}$Si$_{10}$, resulting in weaker changes of the optical properties. For Er$_{5}$Ir$_{4}$Si$_{10}$ a broad shoulder is observed in the specific heat at the lower transition. At the upper transition there is a peak, which is a factor of two smaller as compared to Lu$_{5}$Ir$_{4}$Si$_{10}$\cite{jung03}. All experimental data therefore indicate that the Fermi surface gappings are smaller in Er$_{5}$Ir$_{4}$Si$_{10}$ as compared to Lu$_{5}$Ir$_{4}$Si$_{10}$.

\paragraph{Lu$_{5}$Ir$_{4}$Si$_{10}$}\label{Lu_sub}
\begin{figure}
\includegraphics[keepaspectratio,clip,width=0.48\textwidth]{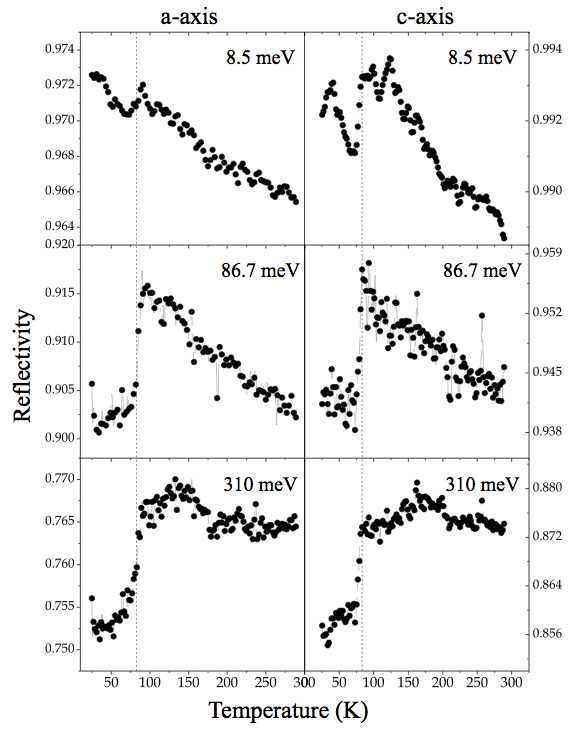}
\caption{\label{RvsT_ac} Temperature dependence of the reflection coefficient of Lu$_{5}$Ir$_{4}$Si$_{10}$ for $E\parallel a$ and $E\parallel c$ at selected frequencies. Both axes show a softened effect for $\hbar \omega < 12.4$ meV $\approx$ 100 cm$^{-1}$.}
\end{figure}
\begin{figure}
\includegraphics[keepaspectratio,clip,width=0.48\textwidth]{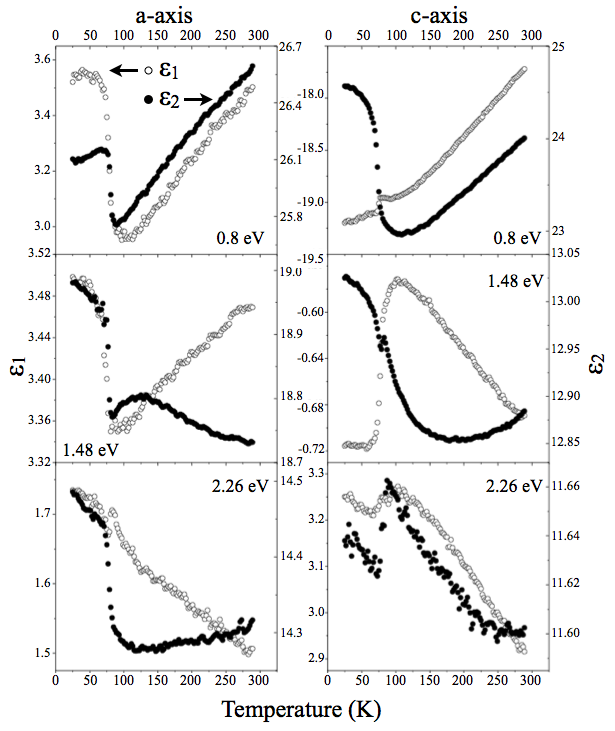}
\caption{\label{Lu_total_epsilon} Temperature dependent dielectric constants of Lu$_{5}$Ir$_{4}$Si$_{10}$. Left hand panels: Real (full circles) and imaginary (open circles) part of the dielectric constant for $E\parallel a$. Right hand panels: The same quantities for $E\parallel c$.}
\end{figure}
From the anisotropy observable in the reflectivity curves, the sample appears
to be more reflecting at low frequency for light polarized along
the \emph{c}-axis, Fig. \ref{fig:IR_Ref}.
The CDW transition is visible in the temperature dependent spectra, Fig.~\ref{RvsT_ac}, as a depletion of the reflectivity below $T_{\textrm{CDW}}$. The drop in reflectivity while cooling the sample is present at all frequencies. We stress that, even though the changes in reflectivity at the transition are extremely clear in the data, the absolute reflectivity variation is less than $0.5\%$ in the low frequency region.
The abrupt change in reflectivity that we observe is located at 78
K, very close to $T_{\textrm{CDW}}$$=83$ K measured using transport techniques. This temperature has been defined as the maximum of $\partial R(\omega_i,T)/\partial T$.

The temperature evolution of the real and
imaginary part of the dielectric function (Fig.~\ref{Lu_total_epsilon}), shows a clear signature of the CDW transition up to
frequencies as high as 2.3 eV. The temperature dependence close to $T_{\textrm{CDW}}$ does not follow the same trend for all the frequencies:  along the \emph{c}-axis (right column of Fig.~\ref{Lu_total_epsilon}) $\varepsilon_2(T)$ increases  upon cooling at $T_{\textrm{CDW}}$ for $\hbar\omega < 1.48$ eV and decreases at higher frequencies. The same behavior is present for $\varepsilon_1(T)$ along both axes but with a different frequency dependence.

For the Er$_{5}$Ir$_{4}$Si$_{10}$ the effects observed at the two transitions (158 K and 53 K) are less than 0.05\% in absolute reflectivity. For this material the influence of the two phase transitions on the optical spectra is too weak to allow a meaningful spectral analysis of the CDW induced features. The CDW induced changes are much larger in the Lu$_{5}$Ir$_{4}$Si$_{10}$ samples, where we observe a step-like change of the optical constants at T$_{CDW}$ at all infrared and optical frequencies and polarizations. The abrupt changes within 10 K around the transition at 83 K justify our approach in some of the subsequent sections of this paper, which is to to characterize the CDW induced change of the optical conductivity by comparing spectra measured 20 K below T$_{CDW}$ with those taken 20 K above T$_{CDW}$.

\section{Strong coupling analysis of the free carrier response}\label{section_lambda}
\begin{figure}
\includegraphics[keepaspectratio,clip,width=0.48\textwidth]{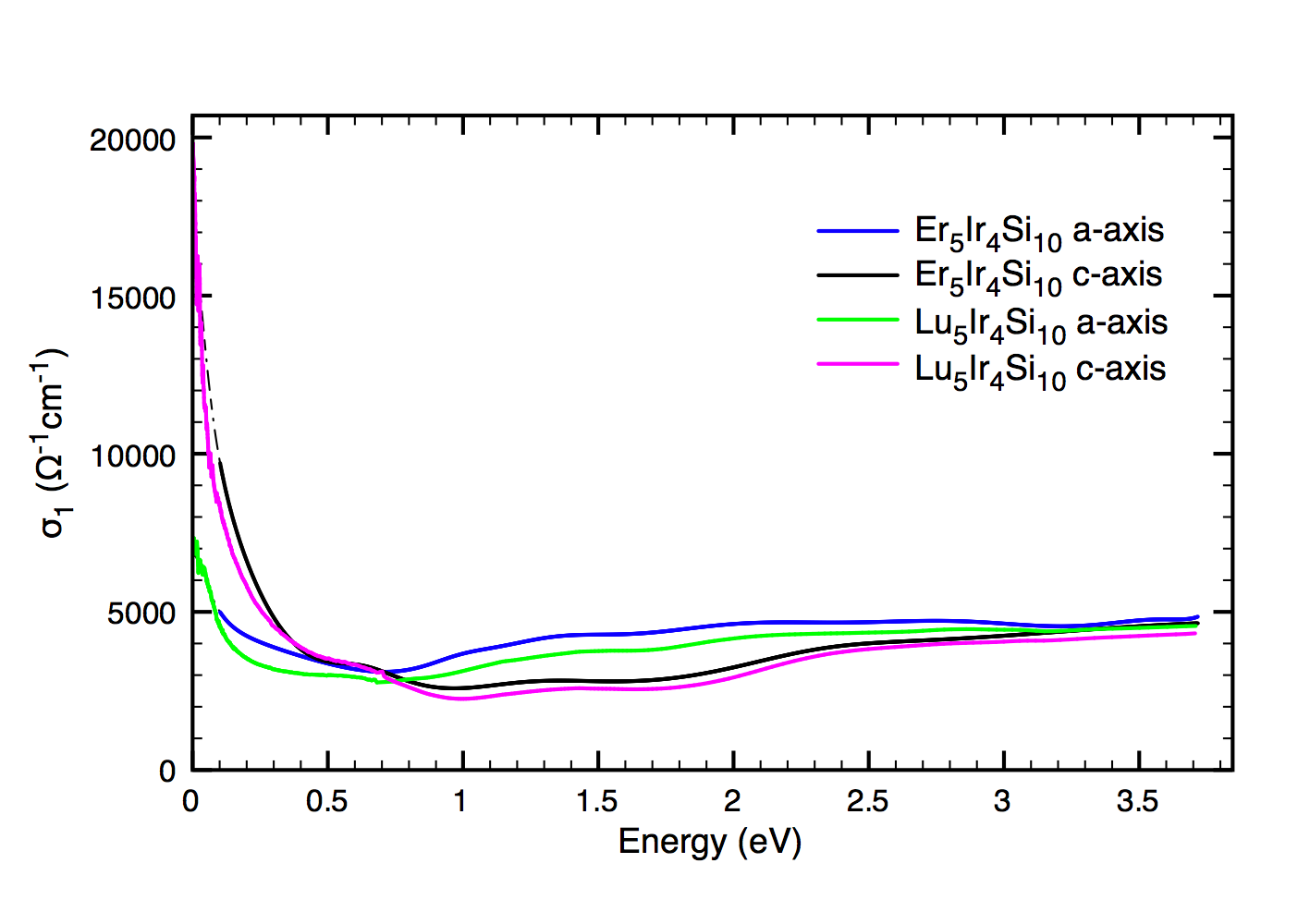}
\caption{\label{Lu_Er_conductivity}
(Color online) Optical conductivity  with $a-$ and $c$-axis polarized light at 290 K of Lu$_{5}$Ir$_{4}$Si$_{10}$ (green for $a$-axis and magenta for $c$-axis) and of Er$_{5}$Ir$_{4}$Si$_{10}$ (blue for $a$-axis and black for $c$-axis).}
\end{figure}
The conductivity curves, displayed in Fig. \ref{Lu_Er_conductivity}, show the expected narrow Drude peak at low frequencies.  In addition the series of minima and maxima for energies larger than 0.5 eV clearly reveals the presence of interband transitions. Later in this section we will see, that the onset of interband transitions in this compound may even be at energies as low as 0.1 eV. It would be very useful to carry out \emph{ab initio} theoretical investigations of the electronic structure of these compounds in relation to the observed optical conductivity. However, a detailed study of the band structure of this material is lacking at the moment. In particular such a study would provide useful information on the energy and optical strength of the interband transitions, and the unrenormalized plasma frequencies.
According to Gruner \cite{gruner} and Kim \cite{kim91}, the CDW has an associated low frequency mode that represents the collective mode of the ordered state. In our experiments the lowest measured
energy was 6.2 meV. For this reason the collective mode
contribution, which lies at even lower energies, was not directly
measured; in our Drude-Lorentz model this contribution is included and hidden in the
the low frequency Drude response.
\begin{figure*}
\includegraphics[keepaspectratio,clip,width=0.8\textwidth]{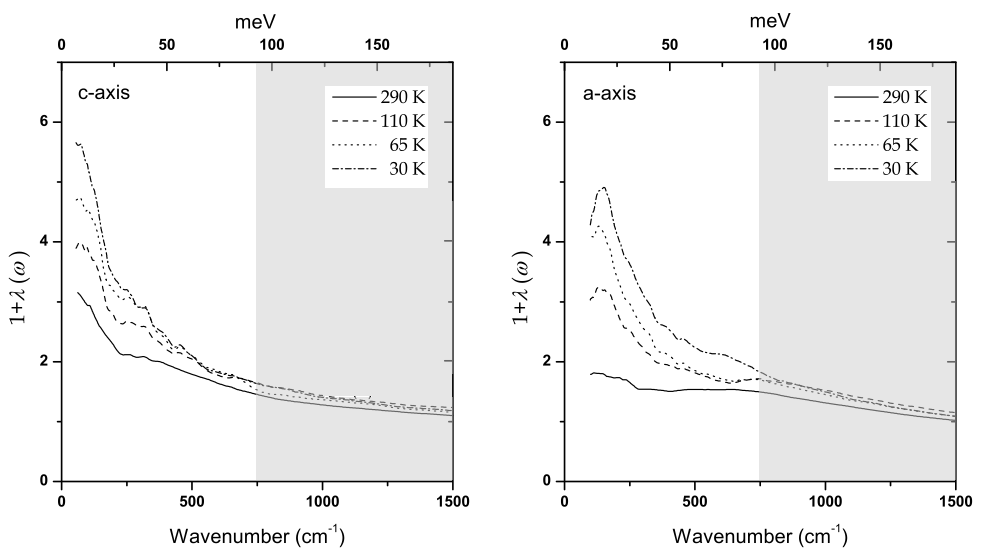}
\caption{\label{Lu_effective_mass_ac} The frequency dependent effective mass $1+\lambda(\omega)$ of Lu$_{5}$Ir$_{4}$Si$_{10}$ for $E
\parallel c$ (left panel) and $E
\parallel ab$ (right panel). Data are shown for the case of 290 K (full line),
110 K (dashed line), 65 K (dotted line) and 30 K (dash-dot line).
Shaded areas are the energy region where the value of $\lambda(\omega)$ may be affected by unaccounted for interband transitions.}
\end{figure*}

The optical conductivity of interacting electrons is described by the extended Drude model\cite{goetze72,allen77}
\begin{equation}\label{eq:extDrude}
    4\pi\sigma(\omega) =
    \frac{i\omega_{\textrm{p}}^{2}}{\omega+\omega\lambda(\omega)+i/\tau(\omega)}.
\end{equation}
where the quantities $1/\tau(\omega)$ and $\lambda(\omega)$
represent, respectively, the frequency dependent scattering rate and
the mass enhancement due to inelastic scattering of the electrons. Their values are obtained from the experimentally measured optical
conductivity via simple inversion formulas:
\begin{equation}\label{eq:OneOverTau}
    \frac{1}{\tau(\omega)} =
    \frac{\omega_{p}^{2}}{4\pi}\textrm{Re}\left(
    \frac{1}{\sigma(\omega)} \right)
\end{equation}
\begin{equation}\label{eq:Lambda}
    1+\lambda(\omega) = \frac{\omega_{p}^{2}}{4\pi}\textrm{Im}\left(
    \frac{-1}{\omega\sigma(\omega)} \right).
\end{equation}
In the above equations the presence of the squared plasma frequency
$\omega_{p}^2 =  4\pi n e^{2}/m^{\textrm{eff}}$ ($n$ is the carrier density
and $m^{\textrm{eff}}$ is the band mass of the free charge carriers) requires the definition of a
cutoff frequency $\omega_{c}$ so that $\omega_{p}$ can be determined
directly by integrating the optical conductivity
\begin{equation}\label{eq:plasma_frequency}
    \omega_{p}(\omega_c)^{2} = 8\int_{0}^{\omega_{c}} \sigma_{1}(\omega)\,
    \ud\omega.
\end{equation}
Usually $\omega_{c}$ is fixed in a frequency region that allows us to
exclude from the integral (\ref{eq:plasma_frequency}) the interband
transition contributions. In our particular case, choosing
$\hbar\omega_{c}$ = 745 (meV) for both polarizations the values of the plasma
energies $\hbar\omega_{p}(\omega_c)$ are 4.3 eV and 3.5 eV for the \emph{c} and \emph{a}-\emph{b}
axes respectively. An important aspect for the following discussion is, that the
temperature variations of the plasma frequencies are negligible ($\sim 0.5$ \%) between 4K and 300 K.
It is important to keep in mind, that the graphs presenting the frequency dependent scattering rate and the effective mass have a consistent physical meaning only in the frequency region where the contributions from interband transition are negligible. The energy regions where we believe that interband transtions are important, will be presented as shaded areas.

In Fig.~\ref{Lu_effective_mass_ac} the effective-mass
$m^{*}(\omega)/m = 1 + \lambda(\omega)$ is shown for
different temperatures above and below the transition. The mass
enhancement is similar for the two optical axis. Above 700 \cmm the
mass enhancement is less pronounced and tends to saturate. Later in this section we will address the cause of this high mass-enhancement, and attribute it to strong coupling to phonons at 200 cm$^{-1}$ ($c$-axis) and 200 cm$^{-1}$ ($a$-axis).
\begin{figure*}
\includegraphics[keepaspectratio,clip,width=0.8\textwidth]{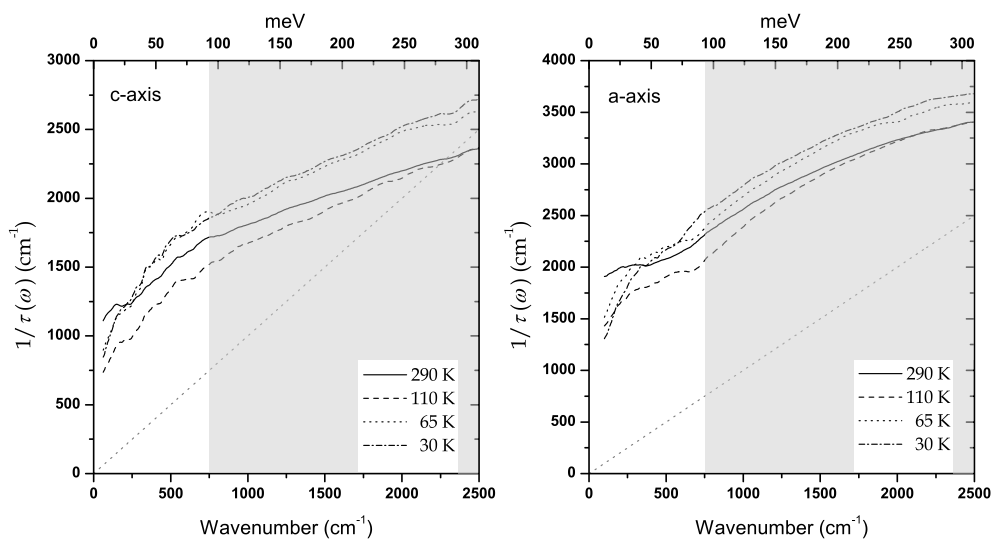}
\caption{\label{Lu_scattering_rate} The frequency dependent scattering rate of Lu$_{5}$Ir$_{4}$Si$_{10}$ $1/\tau(\omega)$ for $E
\parallel c$ (left panel) and $E
\parallel ab$ (right panel). Data are shown for 290 K (full line),
110 K (dashed line), 65 K (dotted line) and 30 K (dash-dot line).
The data shows for both crystal orientations a rigid increase of scattering rate when temperature is lowered below $T_{\textrm{CDW}}$.
Shaded areas are the energy region where the value of $1/\tau(\omega)$ may be affected by unaccounted for interband transitions.}
\end{figure*}
The frequency-dependent scattering rate along the $c$-axis presented in Fig.~\ref{Lu_scattering_rate} reveals a negative curvature, with a linear part above $\omega > 700$ \cmm. This linear section is absent from the \emph{a}-axis curve. However, interband transitions contribute strongly to the optical conductivity along the $a$-axis (see Fig. \ref{Lu_Er_conductivity}), which influences the fine details of $1/\tau(\omega)$, especially at higher frequencies. For both optical axis the behaviour of the scattering rate at low frequencies deviates from the typical electron-electron scattering, $1/\tau \sim \omega^2$, predicted for a standard Fermi liquid as experimentally observed in typical 3D metals \cite{vandereb2001,valla1999}. This peculiarity has been found in several other 2D layered compounds, in particular
2H-TaSe$_{2}$ and 2H-NbSe$_{2}$ \cite{dordevic2001,dordevic2003,valla2000} and in all the high-temperature cuprate superconductors
(HTSC) \cite{basov2002,puchkov1996} where $1/\tau(\omega)$ shows a linear frequency dependence. Another analogy with the above
materials is an abrupt change in the slope of $\tau^{-1}(\omega)$ at low frequencies; in our data this
kink-like feature is visible at around $\Omega_{c} = 700$ \cmm for both crystal orientations.
Even if the frequency dependence of $1/\tau(\omega,T)$ in Lu$_{5}$Ir$_{4}$Si$_{10}$ looks similar to the one observed in HTSC and other quasi-2D layered compounds, its temperature dependence is quite different from these cases: in the under-doped HTSC a progressive suppression of the low frequency scattering rate is observed for temperatures below the pseudogap temperature.
The behavior in Lu$_{5}$Ir$_{4}$Si$_{10}$ is similar to the behavior found along the \emph{ab} plane in 2H-NbSe$_2$ \cite{dordevic2003} where the increase of temperature results in a frequency-independent contribution to the scattering rate, which is progressively added to the $1/\tau(\omega)$ curve.
When the temperature is reduced from 290 K to $T_{\textrm{CDW}}$ we observe a progressive
reduction of the scattering rate, as expected for a typical metallic
system.

In the CDW state part of the Fermi surface is gapped, which has two important consequences:

(i) Part of the free charge carriers is removed from the material, and as a result $\omega_p^2$ which we use to make the conversion from $\sigma(\omega)$ to $1/\tau(\omega)$ and $\lambda(\omega)$ (Eqs. \ref{eq:OneOverTau} and \ref{eq:Lambda}), becomes reduced. In the next subsection we will estimate this reduction to be 15-17\%, but the results shown in Figs. \ref{Lu_effective_mass_ac} and \ref{Lu_scattering_rate} have not been corrected for this reduction. Indeed, when the temperature is lowered below $T_{\textrm{CDW}}$, we observe a sudden increase of $1/\tau$ over the entire frequency region. The change is of the order of
15\% for the \emph{c}-axis and of 10\% for the \emph{a}-axis when comparing the curves relative to $T = 110$ K and $T = 65$ K.

(ii) The spectral weight removed from the free carrier contribution is transferred to the optical transitions across the CDW gap. Since the CDW order results in a reduction of the Brillouin-zone, the transitions across the CDW gap should be regarded as interband transitions in the infrared range. As pointed out above, the contributions of interband transitions should in principle be subtracted from the optical spectra before we calculate $1/\tau(\omega)$ and $\lambda(\omega)$. Since this would be an ambiguous procedure, we have not attempted to do so, and consequently the $1/\tau(\omega)$ and $\lambda(\omega)$ graphs should in principle contain 'spurious' contributions for temperatures below $T_{\textrm{CDW}}$ where the CDW gap has opened. In practice these spurious effects are rather weak in the present case; even the artifacts which are carried over to the experimental $\alpha^2F(\omega)$, which we present below, turn out to be quite subtle.

\begin{figure*}
\includegraphics[keepaspectratio,clip,width=0.8\textwidth]{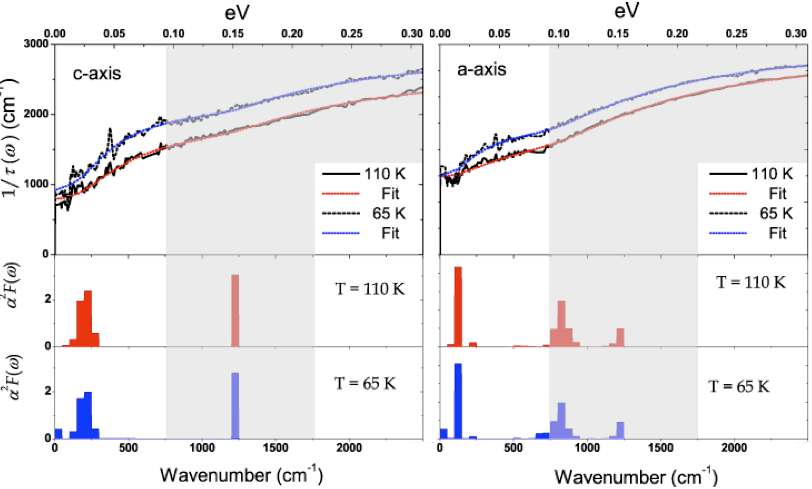}
\caption{\label{ScatteringRateFIT_recolored}
(Color online) The top left and right panels
present a comparison between the experimental scattering rate of Lu$_{5}$Ir$_{4}$Si$_{10}$
measured at 110 K and 65 K and the results of the fitting routine
implemented starting from Eq.~\ref{eq:scattering_rate} for \emph{c}-
and \emph{a}-axis respectively. In the lower panels the resulting
$\alpha^2F(\omega)$ are displayed. The results in the shaded areas should be interpreted with caution: the features in these regions may be artifacts resulting from unaccounted for interband transitions.}
\end{figure*}

The modification to the
electronic properties of metals due to electron-phonon interactions can be described using the Holstein formalism
\cite{holstein64,allen71,dolgov95} in the normal state.
Starting from the single-electron Greens function and introducing the
electron-phonon dependent self-energy, it is possible to derive the
expression for the frequency dependent scattering rate
\begin{eqnarray}\label{eq:scattering_rate}
    \frac{1}{\tau(\omega)} &=& \frac{\pi}{\omega}\int\ud \, \Omega
    \alpha^{2}F(\Omega)
    \cdot \left[
        2\omega\coth \left( \frac{\Omega}{2T} \right) - \right. \nonumber \\
        & &  - (\omega + \Omega)\coth \left( \frac{\omega + \Omega}{2T}
        \right) + \nonumber \\
        & & + \left. (\omega - \Omega)\coth \left( \frac{\omega - \Omega}{2T} \right) +\frac{1}{\tau_0}
    \right]
\end{eqnarray}
where the transport Eliashberg function\cite{eliashberg1960,eliashberg1963}, $\alpha^2F(\omega)$, corresponds to the phonon density
spectrum multiplied by the electron-phonon coupling constant, and $1/\tau_0$ is the residual scattering by impurities.
We used a standard numerical least-squared fitting routine to obtain the spectral function $\alpha^2F(\omega)$, which we model as a series of Einstein modes, for which the variational parameters are the intensities $\lambda_i$ and the frequencies $\omega_{i}$:
\begin{equation}\label{eq:spectral_function}
  \alpha^2 F(\omega) =   \frac{\omega}{2} \sum_{i} \lambda_i \delta(\omega - \omega_i)
\end{equation}
The result of the fitting procedure is presented in Fig.~\ref{ScatteringRateFIT_recolored} where the experimental data at $T = 110$ K and $T = 65$ K are compared with those extracted from the fit. Since the phonon density of states and the electron-phonon coupling are in principle temperature dependent, all temperature dependence observed in $1/\tau(\omega)$ and $\lambda(\omega)$ should be explained by the thermal factors of Eq. \ref{eq:scattering_rate}. An important consistency check on the strong coupling analysis is then to verify the temperature dependence of the $\alpha^2F(\omega)$ output generated by the fit. Indeed, the resulting spectral densities, shown in the lower panels, are remarkably independent of temperature, thus confirming the correctness of the strong coupling model in the present case.

The results of the analysis are presented in Fig.~\ref{ScatteringRateFIT_recolored}: the frequency dependent scattering rate below 500 cm$^{-1}$ originates from a mode centered at 200 cm$^{-1}$ (c-axis), having a width of about 100 cm$^{-1}$. Along the a-axis a peak is seen at 125 cm$^{-1}$, with a width of 50 cm$^{-1}$. As discussed in the introduction, due to the small Drude weight along the $a$-axis interband transitions are particularly clearly visible in the $a$-axis spectra in Fig. \ref{Lu_Er_conductivity}. While with the experimental information at hand we are not able to disentangle interband and intraband transitions unambiguously, both the $a$-axis and $c$-axis optical conductivities have absorption bands as low as 0.5 eV, possibly extending to much lower energy. The conspicuous curvature in the $a$-axis scattering rate shown in Figure \ref{Lu_scattering_rate} is further indication that interband transitions mix in above approximately 65 meV. Hence part of the scattering contributions present in this frequency region is due to interband transitions overlapping with the tail of the Drude peak. The peaks generated by the fitting routine above 65 meV should therefore probably be disregarded disregarded as a physical contribution to $\alpha^2F(\omega)$.

(i) Concentrating now on the frequency range below 500 cm$^{-1}$, we notice first of all, that $\alpha^2F(\omega)$ is remarkably independent on temperature, despite the relative large changes seen in $\sigma_1(\omega)$, $1/\tau(\omega)$, and $\lambda(\omega)$. This is a strong indication, that the latter temperature dependence is correctly described by the thermal factors of the strong coupling formalism used here, which have their origin in the Fermi-Dirac statistics of the electrons at the Fermi-surface, and the Bose-distribution function of the phonons to which they are coupled.

(ii) In the second place we notice the occurrence of a conspicuous low-frequency (25 cm$^{-1}$) mode in $\alpha^2F(\omega)$ in the CDW phase. We believe that this mode is an artifact, which has been generated by attempting to fit gap-related features with a model which does not describe the gap. This implies that, while our analysis correctly provides $1/\tau(\omega)$, $\lambda(\omega)$, and $\alpha^2F(\omega)$ for $T> T_{\textrm{CDW}}$, some caution is necessary for these quantities when $T< T_{\textrm{CDW}}$. Artifacts imported by the incorrect treatment of the partial gapping of the Fermi surface turns out to be subtle: an unphysical zero-frequency mode appears in $\alpha^2F(\omega)$, and $\lambda(\omega\rightarrow 0)$ becomes unrealistically large. These problems do not affect the results for $T > T_{\textrm{CDW}}$. The best estimate is therefore obtained for T=110 K, where along the most conducting direction (the c-axis) $\lambda \approx 2.6$.

\section{The Charge Density Wave gap}\label{section_gap}
\begin{figure}\label{Lu_conductivity_ratio}
\includegraphics[keepaspectratio,clip,width=0.48\textwidth]{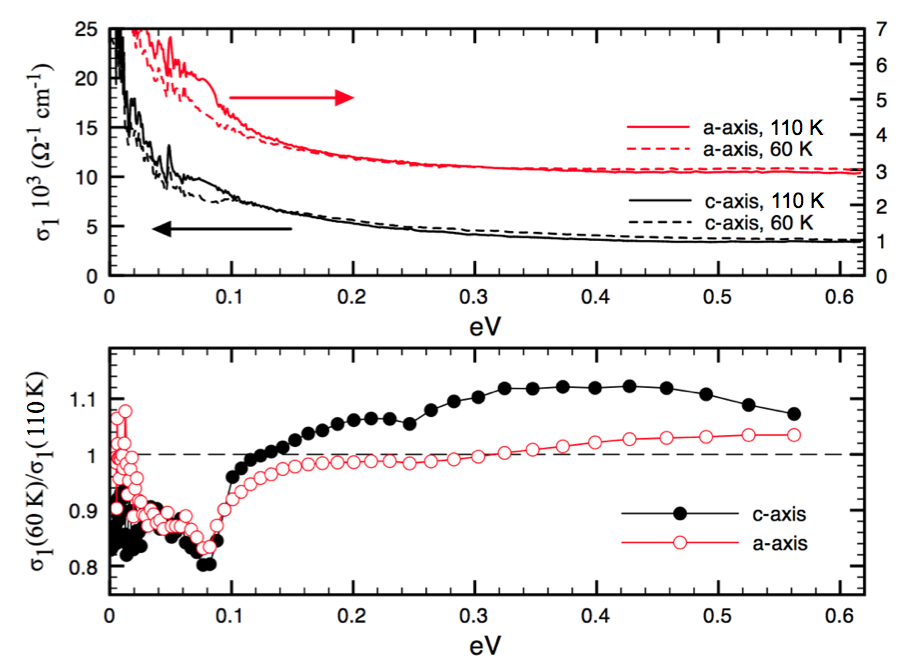}
\caption{\label{Lu_conductivity_ratio}(color online) Top and middle panels: Optical conductivity of Lu$_{5}$Ir$_{4}$Si$_{10}$ along the \emph{a}-axis (red curves) and the \emph{c}-axis (black curves) for $T=60$ K (dashed lines) and $T=110$ K (full line). In the lower panel the ratio $\sigma_{1}(T = 60 K)/\sigma_{1}(T = 110 K)$ reveals a clear reduction of the conductivity below 80 meV for both crystal orientations indicating a partial gapping of the Fermi surface.}
\end{figure}
As pointed out in the previous section, the CDW gap related suppression of $\sigma(\omega)$ imports a subtle and unphysical zero-frequency contribution in the fitted  $\alpha^2F(\omega)$. In order to single out the gap-related features in the optical conductivity spectra, we examine the relative change across the transition temperature. The evolution of $\sigma_{1}$ across $T_{\textrm{CDW}}$ is summarized in the upper panel of Fig.~\ref{Lu_conductivity_ratio}. The low frequency spectra of the optical conductivity are plotted for 60 K (dotted lines) and 100 K (full lines). The optical conductivity of the Lu$_{5}$Ir$_{4}$Si$_{10}$ sample becomes significantly reduced for frequencies below $\sim$ 80 meV when the material enters the CDW state. The most likely reason for this reduction is the opening of a CDW gap on part of the Fermi surface. The fact that the change in the optical conductivity appears in both polarization directions implies, that the Fermi velocities of the states which become gapped are not uniquely along the a or the c-direction.
The ratio $\sigma_{1}(T = 60 K)/\sigma_{1}(T = 100 K)$ (lower panel of Fig.~\ref{Lu_conductivity_ratio}) highlights the CDW induced changes of $\sigma(\omega)$, {\em i.e.} a suppression of $\sigma_1(\omega)$ below $\sim$ 80 meV while at higher energies the conductivity increases. This behavior is in qualitative agreement with a partly gapped Fermi surface, with a momentum dependent gap having a maximum value $2\Delta_{max}\approx$ 80 meV. The partial gapping implies that the Fermi surface is at least strongly warped, or perhaps even entirely 3 dimensional. The decreased conductivity below the CDW gap is compensated by an increase of the conductivity above the gap due to direct optical transitions between occupied and empty bands of the translational symmetry broken CDW state. This gap value corresponds to the ratio $2\Delta_{max}/k_B T_{\textrm{CDW}}\simeq 10$, which is clearly much larger than the weak coupling value $2\Delta=3.52 k_B T_{\textrm{CDW}}$ for an isotropic gap.

A recent infrared study (\Tcdw = 147 K) of polycrystalline Lu$_5$Rh$_4$Si$_{10}$ samples\cite{liu2005} has also revealed a large gap over T$_c$ ratio, $2\Delta/k_BT_c=5.9$, with a Fermi surface gapping of the order of 4\%. Accordingly the Rh polymorph of Lu$_5$Ir$_4$Si$_{10}$ can also be characterized as a strongly coupled CDW.

\section{CDW induced change of free carrier density}\label{section 7}
\begin{figure}
\includegraphics[keepaspectratio,clip,width=0.48\textwidth]{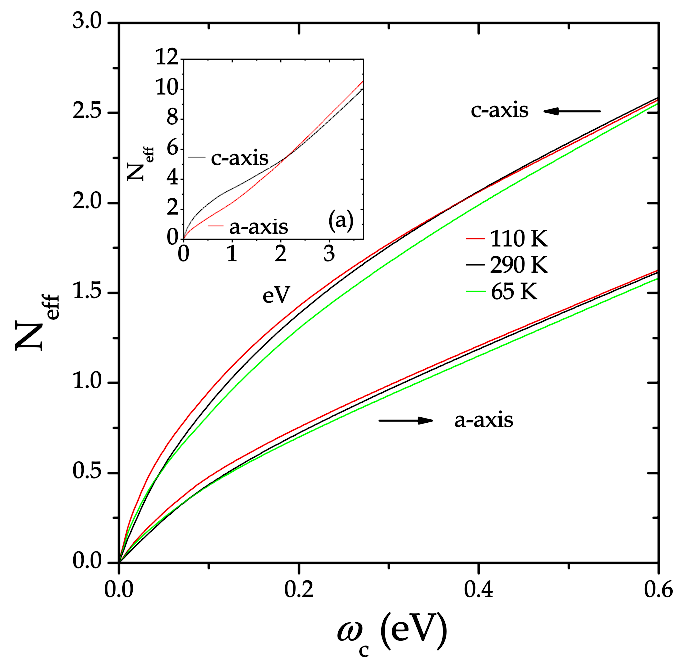}
\caption{\label{Neff2} (Color online) The effective number of carriers of Lu$_{5}$Ir$_{4}$Si$_{10}$ as a function of the cutoff frequency $\omega_c$ for both crystal orientations for $T = 290$ K, $T = 110$ K and $T = 65$ K. In the inset $N_{\textrm{eff}}$ is presented over the full measurement range at room temperature and for both crystal orientations.}
\end{figure}
\begin{figure}
\includegraphics[keepaspectratio,clip,width=0.48\textwidth]{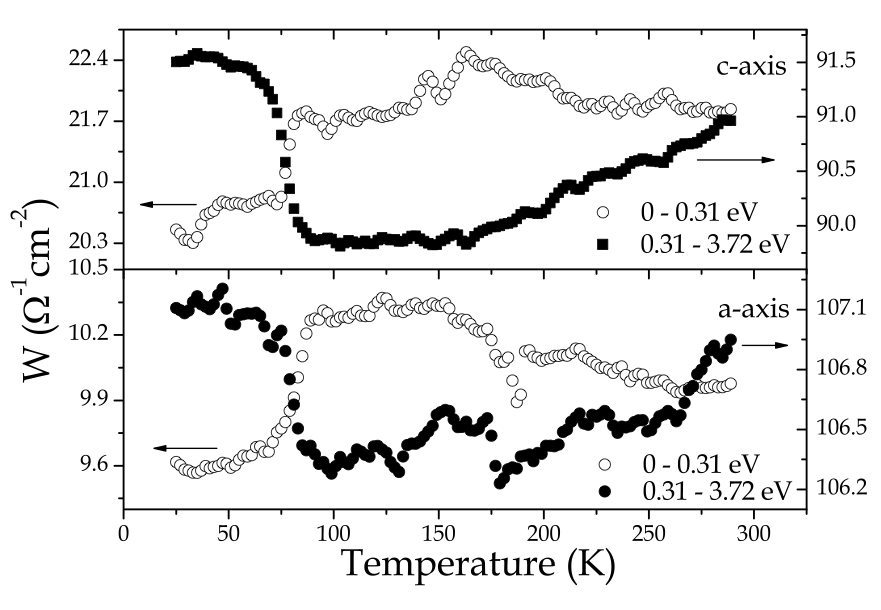}
\caption{\label{Lu_SW} The spectral weight $W(T)$ of Lu$_{5}$Ir$_{4}$Si$_{10}$ along the $a$ (lower panel)
and the \emph{c}-axis (upper panel) calculated using a cutoff frequency
$\omega_{c} = 0.31$ eV. The spectral weight behavior for $\omega>\omega_c$ ($\omega<\omega_c$) is shown with full (empty) markers.}
\end{figure}
\begin{figure}
\includegraphics[keepaspectratio,clip,width=0.48\textwidth]{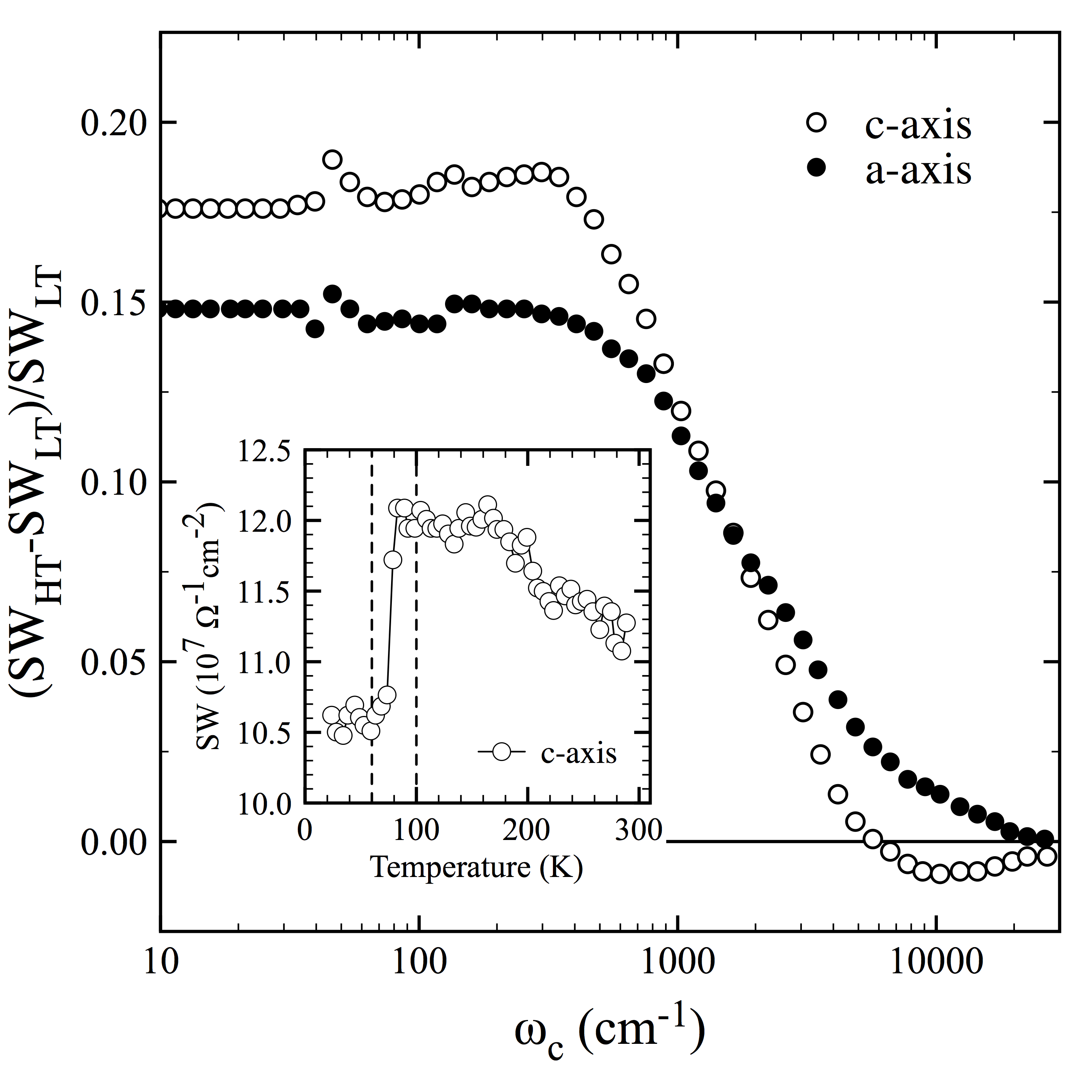}
\caption{\label{SW_ratio} The percentage increase of SW of Lu$_{5}$Ir$_{4}$Si$_{10}$ as
defined in Eq.~\ref{eq:SWpercentage} plotted versus cutoff
frequency. The comparison between c-axis (black-dotted line) and
a-axis (red-dotted line) reveals a saturation region below 45 meV. The recovery of SW extends up to 3 eV for both
optical axis. Inset: Spectral weight integral between 0 and 1000 cm$^{-1}$ as a function of temperature.}
\end{figure}
Information on the temperature dependence of the carriers and the percentage of the CDW gapped part Fermi surface can be obtained from analyzing the optical spectral weight
\begin{equation}\label{eq:spectral_weight}
    W(\omega_c, T) = \frac{\pi e^2}{2m_eV} N_{\textrm{eff}}(\omega_c, T) = \int_{0}^{\omega_c} \sigma_1 (\omega) \, \ud
    \omega.
\end{equation}
In the limit $\omega_c \rightarrow
\infty$ the effective number of charges should equal the
total number of electrons in the system. The dynamics of
$N_{\textrm{eff}}(\omega_c, T)$ crossing a transition temperature quantifies the redistribution of the carriers in the
energy spectrum of the system.
In Fig.~\ref{Neff2} $N_{\textrm{eff}}$ is plotted
versus $\omega_c$ for three different temperatures and for the two
polarization directions for $0 \le \omega_c \le 0.6$ eV; the inset shows the same quantity over
the full measurement range for $T=290$ K. From the temperature dependence of
$N_{\textrm{eff}}$, a noticeable drop in the SW function is clearly visible for both axis cooling the sample
across $T_{\textrm{CDW}}$ indicating a reduction of effective number of carriers. Both along the $a$ and the $c$-axis the low and high temperature
$N_{\textrm{eff}}$ merge at approximately 3 eV, which is equivalent to a full recovery of the SW.

The SW for different temperatures is plotted in the inset of Fig.~\ref{Lu_SW}
for an $\hbar \omega_{c} = 0.31$ eV along both crystal orientations. The graph shows that the low energy SW (represented with open markers) is
decreasing abruptly below $T_{\textrm{CDW}}$. The SW above $\omega_{c}$ (i.~e.~
$\int_{\omega_{c}}^{+\infty}\sigma_{1}(\omega) \, \mathrm{d}\omega$)
has the opposite behavior which suggests that the loss of SW at low
frequencies is compensated by an increase of SW at higher energies.
This behavior is consistent with the CDW model where the opening of a
gap at the Fermi surface causes a spectral weight transfer from the
region within the gap to the energy region of transitions across the CDW gap. The sudden change of spectral weight in a narrow temperature between $\sim$ 80 and 75 K implies that the CDW gap opens almost discontinuously at the phase transition.

Using the information given by the SW, we quantify the percentage of Fermi surface that becomes gapped at the transition by analyzing the temperature variation of the integrated optical spectral weight. For this purpose we define
\begin{equation}\label{eq:SWpercentage}
    \Delta W(\omega_c) = \frac{W(\omega_c, T=110 \, K)-W(\omega_c, T=65 \, K)}{W(\omega_c, T=65 \, K)}.
\end{equation}
This quantity, shown in Fig. \ref{SW_ratio}, vanishes for $\hbar\omega_c>3$ eV, because all low energy spectral weight removed by the gap is recovered at 3 eV. For energies larger than the Drude width and smaller than the gap $\Delta W(\omega_c)$ corresponds to the percentage increase of carriers by closing the gap. The temperature dependence of SW is shown
in the inset of Fig.~\ref{SW_ratio} for $\hbar\omega_c = 124$ meV (1000 \cmm):
a clear drop in $SW$ is visible at $T_{\textrm{CDW}}$ meaning that a certain
amount of carriers is removed from the frequency region below
$\omega_c$ and transferred at higher frequencies. We observed the same behavior along the \emph{a}-axis. We see that all optical data imply a gap induced removal of about 15-17\% of the free carrier spectral weight. Based on the increase of the superconducting temperature at the critical pressure of 21 kbar, Shelton {\em et al.}\cite{shelton1986} estimated that 36 \% of Fermi surface which becomes gapped by the CDW. In view of the indirectness of this method, the latter value is not inconsistent with the 15-17 \% change in free charge carrier density obtained from the optical spectral weight analysis.

\section{Conclusions}

We have studied the optical response of two members of the RE$_{5}$Ir$_{4}$Si$_{10}$ family: Lu$_{5}$Ir$_{4}$Si$_{10}$ and Er$_{5}$Ir$_{4}$Si$_{10}$. %
The experimentally determined temperature dependent optical constants showed, for both samples, sudden, first order like, changes
at $T_{\textrm{CDW}}$. This behaviour has been observed also in other experiments \cite{kuo01,becker99} and indicates the absence of fluctuations which characterize a CDW system of weakly coupled one-dimensional chains, and the presence of a positive feedback mechanism in the CDW order, probably resulting from a coupling of the CDW order parameter to a static periodic modulation of the lattice parameter. This observation is compatible with the observation, that the entropy change at the CDW transition in Lu$_{5}$Ir$_{4}$Si$_{10}$ is dominated by strong phonon contributions. Both samples of the RE$_5$Ir$_4$Si$_{10}$ family have optical characteristics of an 3D material with modest anisotropy of the plasma frequencies and effective mass of the conduction electrons, although the observed CDW transition clearly indicates that some ($\sim 15\%$) portions of the Fermi surface are nested below $T_{\textrm{CDW}}$.
The Eliashberg function $\alpha^2F(\omega)$ was obtained from a fit for both crystallographic directions, showing clear evidence for strong coupling ($\lambda\approx 3$) of the electrons to phonons centered at 30 meV.
From a careful analysis of the free carrier spectral weight, we estimated the amplitude of the CDW gap to be 80 meV, implying $2\Delta/k_{B}T_{\textrm{CDW}} \simeq 10$.

\section{Acknowledgments}
We gratefully acknowledge stimulating discussions with F.
Marsiglio and S. Ramakrishnan. This work is supported by the Swiss National Science Foundation through Grant
No. 200020-113293 and the National Center of Competence in
Research (NCCR) "Materials with Novel Electronic
Properties-MaNEP."

\end{document}